\documentclass[numsec,webpdf,contemporary,large,namedate]{oup-authoring-template}

\usepackage{amsmath}
\usepackage{amsthm}
\usepackage{tabularx}
\usepackage{booktabs}
\usepackage{longtable}
\usepackage{tcolorbox}
\onecolumn
\usepackage[onehalfspacing]{setspace}

\graphicspath{{Fig/}}

\theoremstyle{thmstyleone}%
\newtheorem{theorem}{Theorem}
\newtheorem{proposition}[theorem]{Proposition}%
\theoremstyle{thmstyletwo}%
\newtheorem{condition}{Condition}%
\newtheorem{corollary}{Corollary}%
\theoremstyle{thmstylethree}%

\usepackage[fontsize=12pt]{fontsize}

\begin{document}

\journaltitle{}
\DOI{}
\copyrightyear{2026}
\pubyear{2026}
\access{}
\appnotes{Article}

\firstpage{1}


\title[Embracing Spillover]{Embracing Spillover: Spatial Effects in Experiments}

\author[1,$\ast$]{Samuel I. Watson}
\author[2]{Penny Hancock}
\author[1]{Jack Fraser-Govil}
\author[2]{Kieran DeWalt}
\author[3]{Thomas A. Smith}

\authormark{S.I. Watson et al.}

\address[1]{\orgdiv{Department of Applied Health Sceinces}, \orgname{University of Birmingham}, \orgaddress{\street{Birmingham},  \country{United Kingdom}}}
\address[2]{\orgdiv{Department of Infectious Disease Epidemiology}, \orgname{Imperial College, London}, \orgaddress{\country{United Kingdom}}}
\address[3]{\orgname{Swiss Tropical and Public Health Institute}, \orgaddress{\country{Switzerland}}}

\corresp[$\ast$]{Address for correspondence. Samuel I. Watson, University of Birmingham, Birmingham, B15 2TT, United Kingdom. \href{Email:s.i.watson@bham.ac.uk}{s.i.watson@bham.ac.uk}}




\abstract{Interventions delivered in space generate effects that spill over between experimental units. We develop a framework for spatial experiments in which causal estimands, inclduing direct, indirect, total, and dose--response effects, are linear functionals of a spillover kernel. Under cluster randomisation with a fixed exposure set size, the data identify the shape of the kernel but not its level: the level is aliased with the intercept and direct effect, at any sample size and under any outcome model. Every estimator therefore rests on an anchoring assumption that fixes the level. We derive an exact decomposition of the bias of any anchored estimator into a level term, a shape error from kernel misspecification, and a leakage term absorbed by the realised geometry, each computable from the design before data collection. A single scalar, the level leverage, gives each estimand's exposure to the unidentified level, the exact level bias, and the variance cost of estimating the level instead. The conventional cluster-trial analysis is the special case of an implicit anchor, with contamination bias in closed form. Existing approaches, including identification through Bernoulli randomisation, elicited bounds on interference decay, and assumed compact support, are, within the linear exposure mapping, anchoring choices in this framework. We give a taxonomy of design augmentations that purchase the level and a criterion for when to augment and when to anchor.
}

\maketitle

\section{Introduction}
\label{sec:intro}

\subsection{Background}
Many interventions in health and medicine have inherently spatial effects. Archetypal examples in the health space are interventions that prevent the transmission of vector-borne illnesses, such as providing insecticide treated bed nets \citep{Binka1998, Mosha2024}, or new technologies like releasing genetically modified mosquitos \citep{Habtewold2026}. Typically the experimental design used to evaluate the effectiveness of these interventions is the cluster randomised trial (CRT). CRTs involve groups of participants, which are randomly allocated to treatment and control conditions \citep{Eldridge2012}. These designs are typically used when the intervention affects the whole or part of the group, or when intervening on any member of the group affects others within the group. In these cases, the standard individual randomized controlled trial design would generate a biased estimator of an average treatment effect (ATE) due to `spillover' between treated and control individuals. As such the whole group is randomized together. 

The spatial nature of the interventions in some CRTs can lead to a violation of core identifying assumptions. For cluster randomized, and other types of trials, an identifying assumption is the stable unit treatment value assumption (SUTVA), which states that an individual's or cluster's outcome is only determined by their treatment status and not by the allocation of any other unit. However, if, by reducing the risk of, for example, malaria in treated clusters, we also reduce the risk in nearby control clusters we violate SUTVA and bias the treatment effect estimator. To try to ensure SUTVA is satisfied in trials with a spatial setting, trialists often aim to space clusters far enough apart to prevent spillover. A common approach is the `fried egg' design, which includes a buffer zone around each cluster excluded from analyses \citep{Hayes2009}. Sufficiently spacing clusters is not always feasible though and some recent large cluster trials have acknowledged the effect of spillover. For example, a trial of Wolbachia-infected mosquito deployments to reduce the risk of dengue fever, acknowledged spillover and computed an exposure index for participants who commuted between clusters, and carried out a supplementary analysis allowing for spillover caused by human movement \citep{Utarini2021}. 

In this article, we argue that the traditional spaced CRT design has potential flaws when the intervention exhibits spatial effects. First, the spacing between clusters may make the assumption of no between cluster spillover untestable. Second, estimated own-trial ATEs from traditional spaced CRTs are not transportable to new geometries. And third, estimands beyond the ATE are often not identifiable from the data. 

\subsection{Spatial effects in experiments}
There are several mechanisms by which an intervention can exert its effects spatially. First, for some interventions effects are generated from a source (e.g. facility, release site, pollution origin). Everyone has a continuous exposure to the intervention that is a function of distance to source either through diffusion mechanism or a propensity to travel to the site. In these cases, there may be interest in estimating a dose-response function. Second, in a `direct/indirect' framework treatments are assigned to discrete units (e.g. households or individuals) and the outcome for any unit depends on its own treatment status and those of its neighbours. The indirect effects are also a continuous function of distance. Indeed, this article argues that the two frameworks are unified by a common distance kernel that is assumed to be transportable. 

Despite the range of potential spatial mechanisms of many interventions, incorporation of spatial models in experimental design and analysis is rare \citep{Jarvis2017}. Three recent papers address spatially structured estimands directly, and each resolves the same underlying problem by a different route. \citet{Wang2025} show that under Bernoulli assignment, randomisation alone nonparametrically identifies the marginal effect of activating treatment at a given distance: both arms of the switch have positive probability by design, and the estimand references the design's own ambient exposure distribution rather than a counterfactual the design cannot reach. \citet{Leung2025} instead restricts the reach of interference, eliciting a bound on its spatial decay and pricing the residual bias into estimator choice, cluster construction, and bias-aware intervals. \citet{Watson2025} assumes compact support of the spatial effect kernel outright, which identifies dose--response estimands under a range of randomisation schemes, but does not examine the consequences of kernel misspecification or of the support assumption failing. Many of the quantities trialists report resist all three routes as stated: the direct effect of treatment, the total effect on a treated cluster, and the spillover-distance curve referenced to the absence of intervention depend on outcomes under a counterfactual with no treatment anywhere, an allocation to which cluster-randomised designs assign zero probability, and which their fixed margins prevent the data from approaching. We show in this article that this dependence is the source of a structural non-identifiability that randomisation cannot resolve, characterise exactly which estimands are exposed to it, and quantify what every estimator must assume, implicitly or explicitly, to report them.

Other recent methodological work in this area has focused on cluster selection and cluster design in the context of a bias-variance trade-off for the difference in means estimator, rather than explicitly incorporating the spatial dependence structure for the experimental design and analysis \citep{Leung2025}. There have been several re-analyses of prior trials that demonstrate spatially varying effects of interventions as functions of distance from intervention locations (e.g. \citet{Jarvis2019, AnayaIzquierdo2021, Multerer2021}). \citet{Watson2021} considered efficient cluster design under spatial correlation. Other work in this area focuses on approaches to minimising spillover, such as through restricted randomisation \citep{McCann2018}.

In other settings, particularly network-based inference, both direct and indirect effects of treatments are considered \citep{Forastiere2021, Leung2022}. \citet{Hudgens2008}'s partial interference framework considers interference \textit{within} groups and the estimation of direct and indirect effects, for example. Vaccine cluster randomised trials also consider direct and indirect effects with different levels of vaccine exposure within clusters \citep{Kilpatrick2020}. Our framework considers that the spillover effect is mediated non-linearly through some smooth function of distance in a relevant metric, which requires extending the methods for direct and indirect effects to a spatial domain.

\subsection{Outline}
Our central claim is that spacing clusters and intervention sites can make spillover unverifiable from the trial outcome data over the distances that matter. Whilst trial data can potentially identify the shape of the spillover kernel, the level of spillover is unindentified, and so the conventional remedy of separating the clusters cannot guarantee that between-cluster spillover is removed, but instead is precisely the design which renders this assumption unfalsifiable. When residual spillover is scientifically plausible, design augmentation can convert an untestable assumption into an estimable component.

We present the mathematical arguments in four steps: first, we set up a general framework for spatial estimands; second, we consider the identifiability of these estimands; third, we develop an analysis of estimation of the estimands; and fourth, we present quantitative measures of the issues we identify. The analysis is supplemented with a short simulation-based study.

\section{Estimands}
\subsection{Setup and potential outcomes}
 We assume a spatial, geographic setting. Let $N$ individuals be located at $s_1,...s_N$ in a compact domain $\mathcal{D}$. We assume a Euclidean plane $\mathcal{D} \subset \mathbb{R}^2$. We let $k(i) \in \{1, ... , K \}$ denote the cluster of individual $i$. For simplicity we assume there are $n$ individuals per cluster so that $Kn = N$, but we note that $n$ could be one and the framework does not explicitly require a clustered design. The treatment is assigned at the cluster level $\mathbf{z} \in \{0,1\}^K$ and is drawn under a balanced randomisation scheme $\mathbf{z} \in \mathcal{Z}$.  We notate $z_{k(i)}$ as the treatment status of cluster $k$ and $\tilde{\mathbf z}$ as the unit level vector of treatment effects. We define a `source configuration' for the locations of the indirect spatial effects $\mathcal{T} = \{t_1, ..., t_M \} \subset \mathcal{D}$ with a cluster mapping $\iota: \mathcal{T} \to \{1,...,K\}$ which carries treatment $z_{\iota(t)}$. The sources could be other treated individuals or intervention sources as in \citet{Watson2025}. Each individual has an exposure set $\mathcal{T}_i = \mathcal{A}_i \setminus E_i$, where $E_i \subset \mathcal T$ is an exclusion rule that removes a fixed number $q_i \geq 0$ of own-cluster sources, and $\mathcal A_i \subseteq \mathcal T$ are sources eligible to affect $i$, which may be smaller than $\mathcal{T}$ through exclusions, design restrictions, or otherwise. The distance metric is $d(s_i,t)$ where $t \in \mathcal{T}_i$. The distance may be Euclidean but our analysis is metric and domain agnostic. The maximum pairwise distance between two locations is $D$.

\subsection{Source counts and exposure measures}
We introduce counting notation for the intervention sources. Let $A_{ik} = |\{t \in \mathcal A_i : \iota(t) = k\}|$ be the number of unit $i$'s excluded sources belonging to cluster $k$, and $M_k := |\{t \in \mathcal T : \iota(t) = k\}|$ the number of sources belonging to cluster $k$, with
\begin{equation*}
    M(\mathbf z) := \sum_{t \in \mathcal T} z_{\iota(t)}     = \sum_{k=1}^K M_k z_k
\end{equation*}
the total treated-source count under allocation $\mathbf z$. When cluster source counts are equal we write $m_0$ for their common value, so $M_k = m_0$ for all $k$; the margin constant $M_0$ below is unrelated to the cluster counts $M_k$. Three design properties recur throughout:
\begin{enumerate}
\item[(H1)] \emph{(Global exposure)} every source lies within the modelled range of every unit: $\mathcal T_i = \mathcal T \setminus \mathcal A_i$ for all $i$;
\item[(H2)] \emph{(Homogeneous exclusion)} $\mathcal A_i$ contains only sources in cluster $k(i)$, with $|\mathcal A_i| = q$ for all $i$;
\item[(H3)] \emph{(Fixed margin)} $M(\mathbf z) = M_0$ for all $\mathbf z \in \mathcal Z$.
\end{enumerate}
Typically $q = 1$, corresponding to excluding the focal unit's own source from the spillover sum so that own treatment is captured by the direct-effect parameter rather than double-counted as indirect exposure, for $q = 0$ for interventions whose sources are separate from the
beneficiaries.

Estimands and spillover summaries are expressed through measures on distance. Let $w$ be a finite signed measure on $[0,D]$, with $D$ the maximum pairwise distance, and $c \in \{0,1\}$ a direct-effect indicator, and define
\begin{equation*}
    \langle f, w \rangle := \int_0^{D} f(u)\, w(du), \qquad     m(w) := w([0,D]), \qquad     \ell(w,c) := m(w) + qc.
\end{equation*}
We call $m(w)$ the level mass and $\ell(w,c)$ the level leverage of the estimand $(w,c)$: it counts, per unit, the exposure pairs through which the estimand references the unexposed counterfactual. Writing $\Delta_d$ for a unit point mass at distance $d$, so $\langle f, \Delta_d \rangle = f(d)$, the empirical exposure measures are
\begin{equation*}
    w_W := \frac{1}{N}\sum_{i=1}^N     \sum_{\substack{t \in \mathcal T_i \\ \iota(t) = k(i)}}     \Delta_{d(s_i,t)}, \qquad     w_B := \frac{1}{N}\sum_{i=1}^N     \sum_{\substack{t \in \mathcal T_i \\ \iota(t) \neq k(i)}}
    \Delta_{d(s_i,t)},
\end{equation*}
so that $m(w_W)$ and $m(w_B)$ are the average per-unit within- and between-cluster source counts. Under (H1)--(H2) with equal cluster source counts and $q = 1$, $m(w_W) = m_0 - 1$ and $m(w_B) = (K-1)m_0$; the fixed margin (H3) plays no role in these counts. The anchor measure $h$ is the limiting density of the exposure distances, the normalised limit of $w_W + w_B$ under Condition~C3(ii); its within- and between-cluster components $h_w$ and $h_b$ are defined where they are first used, at the estimands table.

\subsection{Finite-population causal estimands}
For individual $i$, let $Y_i(x,\mathbf a_i)$ denote the potential outcome when the individual's own direct treatment status is $x\in{0,1}$ and the active-source configuration in their exposure set $\mathcal T_i$ is $\mathbf a_i=(a_{it}:t\in\mathcal T_i)$ for $a_{it}\in\{0,1\}$. Under the realised allocation the arguments are $x = z_{k(i)}$ and $a_{it} = z_{\iota(t)}$, and we write $Y_i(\mathbf z)$ for the resulting potential outcome. The configurations in
Table~\ref{tab:finite_population_estimands} are not of this form.

Let $\mathbf 0_i$ denote the source configuration in which all sources in $\mathcal T_i$ are inactive. Let $\mathbf a_i^W$ denote the configuration in which all own-cluster sources in $\mathcal T_i$ are active and all other sources are inactive $a_{it}^W=\mathbf 1{\iota(t)=k(i)}$. Let $\mathbf a_i^B$ similarly denote the configuration in which all between-cluster sources in $\mathcal T_i$ are active and $\mathbf a_i^{WB}$ denote the configuration in which all sources in $\mathcal T_i$ are active. For dose-response estimands, let $\mathbf a_i^{d}$ denote a hypothetical source configuration with exactly one active source at distance $d$ from individual $i$ and no other active sources. 

Table \ref{tab:finite_population_estimands} shows different estimands within a fixed-margin, homogeneous exclusion framework, including a standard ATE under a universal roll-out, but also including dose-response  \citet{Watson2025, Wang2025}, and a spatial direct and indirect effect. These estimands also extend existing approaches to direct and indirect effects \citep{Forastiere2021, Leung2022}, two-level effects \citep{Vanderweele2013}, and partial interference \citep{Hudgens2008}. 

\begin{table}[]
\small
\centering
\begin{tabularx}{\textwidth}{cX X}
\toprule
Estimand & Finite-population definition & Interpretation \\
\midrule
$\tau_{DE}$ & $\displaystyle \frac{1}{N}\sum_{i=1}^N \{ Y_i(1,\mathbf 0_i)-Y_i(0,\mathbf 0_i) \}$ & Own direct treatment effect when all exposure sources are inactive.   \\
$\tau_W$ & $\displaystyle \frac{1}{N}\sum_{i=1}^N \{ Y_i(0,\mathbf a_i^W)-Y_i(0,\mathbf 0_i) \}$ & Indirect effect of activating own-cluster sources, holding own direct treatment inactive.  \\
$\tau_B$ & $\displaystyle \frac{1}{N}\sum_{i=1}^N \{ Y_i(0,\mathbf a_i^B)-Y_i(0,\mathbf 0_i) \}$ & Indirect effect of activating between-cluster sources, holding own direct treatment inactive. \\
$\tau_{\mathrm{cluster}}$ & $\displaystyle \frac{1}{N}\sum_{i=1}^N \{ Y_i(1,\mathbf a_i^W)-Y_i(0,\mathbf 0_i) \} $ & Effect of treating the individual's own cluster, including own direct treatment and own-cluster source exposure.  \\
$ATE_{\mathrm{trial}}$ & $\displaystyle \frac{1}{N}\sum_{i=1}^N \{ Y_i(1,\mathbf a_i^{WB})-Y_i(0,\mathbf 0_i)\} $ & Effect of universal rollout in the trial geometry.  \\
$\theta(d)$ & $\displaystyle \frac{1}{N}\sum_{i=1}^N \{ Y_i(0,\mathbf a_i^d)-Y_i(0,\mathbf 0_i) \} $ & Effect of activating a single source at distance $d$, with own treatment inactive and all other sources inactive. \\
$\theta(d)-\theta(d')$ & $\displaystyle \frac{1}{N}\sum_{i=1}^N \{ Y_i(0,\mathbf a_i^d)-Y_i(0,\mathbf a_i^{d'}) \} $ & Contrast between a source at distance $d$ and a source at distance $d'$. \\
$\bar\theta_{h_s}$ & $\displaystyle \int \left[ \frac{1}{N}\sum_{i=1}^N \{ Y_i(0,\mathbf a_i^u) -Y_i(0,\mathbf 0_i)\}\right]h_s(u),du$ & Mean effect of placing one new source, where the distance from individuals to that source has density $h_s$. \\
\bottomrule
\end{tabularx}
\caption{Finite-population causal estimands and their model-implied forms under the additive spatial response model. The definitions do not require the additive model; the final column gives the corresponding expression when $Y_i(x,\mathbf a_i)=\alpha+\tau x+\sum_{t\in\mathcal T_i}\phi(d_{i,t})a_{it}+\varepsilon_i$.}
\label{tab:finite_population_estimands}
\end{table}

\subsection{Estimands as linear functionals}
We specify an additive model for the potential outcomes:
 \begin{equation}
 \label{eq:spatmod}
     Y_i(\mathbf{z}) = \alpha + \tau z_{k(i)} + S[\phi]_i(\mathbf{z}) + \varepsilon_i, \hspace{1cm} \varepsilon \sim N(0,\Sigma)
 \end{equation}
 with spillover exposure $S[\phi]_i(\mathbf{z}):= \sum_{t \in \mathcal{T}_i} \phi(d_{i,t})z_{\iota(t)}$, where $\phi : [0, D] \to \mathbb{R}$ is the true spillover kernel.  The spillover term is an exposure mapping \citep{Aronow2017, Svje2024} linear in treated sources with a nonparametric distance kernel, so misspecification of the kernel's shape is accommodated throughout, while nonlinearity in the exposure itself is not. 

Every causal quantity we consider is an estimand of the form:
\begin{equation}
\label{eq:linearf}
    \theta := c \tau + \langle \phi, w\rangle
\end{equation}
Table $\ref{tab:estimands}$ shows the different estimands and the respective values of the components of (\ref{eq:linearf}). Time enters the framework as columns of $X_0$ rather than as new structure. We develop the consequences, including the level-aliasing of standard time-adjusted staggered roll out models, in \S\ref{sec:temporal}, and keep the kernel static throughout.

\begin{table}[]
    \centering
    \begin{tabularx}{\textwidth}{c|Xcccc}
    \toprule
    Estimand & Model form & $c$ & $w$ & $\ell(w,c)$ & $\ell(w,c)$* \\
    \midrule
    $\tau_{DE}$ & $\displaystyle \tau$ & 1 & 0 & $q$ & $1$  \\
    $\tau_W$ & $\displaystyle \langle\phi, w_W\rangle $ & 0 & $w_W$ & $m(w_W)$ & $m_0 -1$ \\
    $\tau_B$ & $\displaystyle \langle\phi_w, w_B\rangle$  & 0 & $w_B$ & $m(w_B)$ & $(K-1)m_0$ \\
    $\tau_{\text{cluster}}$ & $\displaystyle \tau+ \langle\phi, w_W\rangle$ & 1 & $w_W$ & $m(w_W) + q$ &  $m_0$\\
    $ATE_{trial}$ & $\displaystyle \tau+\langle\phi, w_{W}+w_B\rangle$ & 1 & $w_W + w_B$ & $m(w_W) + m(w_B) + q$ & $Km_0$ \\
    $\phi(d) - \phi(d')$ & $\displaystyle {\langle \phi, \Delta_d- \Delta_{d'}\rangle}$ & 0 & $\Delta_d - \Delta_{d'}$ & 0 & 0 \\
    $\phi(d)$ & $\displaystyle \langle \phi, \Delta_d\rangle $ & 0 & $\Delta_d$ & 1 & 1 \\
    $\bar{\phi}_{h_s}$ & $\displaystyle \langle \phi, h_s\rangle $ & 0 & $h_s$ & 1 & 1 \\
         \bottomrule
    \end{tabularx}
    \caption{Estimands expressed as linear functionals. *values of $\ell(w,c)$ under the fixed-margin, homogeneous exclusion case with $q=1$}
    \label{tab:estimands}
\end{table}

\subsection{Transportability}
The parameters $(\tau, \phi)$ are properties of the intervention mechanism. If the mechanistic kernel is stable across settings after choosing the appropriate distance metric and covariates, then geometry-specific effects can be transported by changing $w$. Effects in the target are obtained by evaluating the functional at new $w$ including distributions $(h_w, h_b)$ and a new configuration $\mathcal{T}'$. 

\section{Identifiability}
A key structural result is the level non-identifiability of the spillover kernel. Proofs of this and the subsequent statements are provided in the Supplementary Material.

\begin{proposition}[Level aliasing]
\label{prop:level_aliasing}
Write $\boldsymbol\eta(\mathbf z) = \alpha\mathbf 1 + \tau\tilde{\mathbf z} + S[\phi](\mathbf z)$ for the linear predictor, $X_0(\mathbf z) = [\mathbf 1, \tilde{\mathbf z}]$, and $s_0(\mathbf z) = S[\mathbf 1](\mathbf z)$. \begin{enumerate}
\item[(i)] For a realised allocation $\mathbf z$, the level of $\phi$ is aliased with $(\alpha, \tau)$ if and only if $s_0(\mathbf z) \in \mathrm{col}\{X_0(\mathbf z)\}$: writing $s_0(\mathbf z) = X_0(\mathbf z)b$, the triples $(\alpha, \tau, \phi)$ and $(\alpha - ab_0,\ \tau - ab_1,\ \phi + a)$ induce the same $\boldsymbol\eta(\mathbf z)$ for every $a \in \mathbb R$.
\item[(ii)] If every source lies within the modelled range of every unit (H1) and each unit excludes exactly $q$ own-cluster sources (H2), then $s_0(\mathbf z) = M(\mathbf z)\mathbf 1 - q\tilde{\mathbf z}$, so the criterion holds at every allocation, whatever the randomisation scheme. If additionally the source margin is fixed, $M(\mathbf z) = M_0$ on $\mathcal Z$ (H3), the flat direction is common to all allocations:
\begin{equation*}
(\alpha,\ \tau,\ \phi) \;\longmapsto\; (\alpha - aM_0,\ \tau + aq,\ \phi + a), \qquad a \in \mathbb R.
\end{equation*}
\item[(iii)] Under (H1)--(H3), moving along the flat direction changes $\theta(w,c) = c\tau + \langle\phi, w\rangle$ by $a\,\ell(w,c)$, where $\ell(w,c) := m(w) + qc$. Hence $\theta(w,c)$ is unidentified without an anchor (\S\ref{sec:anchors}) whenever $\ell(w,c) \neq 0$, while estimands with $\ell(w,c) = 0$ are invariant to the unidentified level.
\item[(iv)] Parts (i)--(iii) hold throughout the class of models in which the law of $\mathbf y$ given $\mathbf z$ depends on $(\alpha, \tau, \phi)$ only through $\boldsymbol\eta(\mathbf z)$: the linear model with any covariance $\Sigma$ free of $(\alpha,\tau,\phi)$, generalised linear models with $\mathbb E(y_i \mid \mathbf z) = g^{-1}\{\eta_i(\mathbf z)\}$, and mixed models whose conditional law given random effects depends on the parameters only through $\boldsymbol\eta(\mathbf z) + Z\mathbf b$. In this class the flat direction leaves the likelihood invariant at every $\mathbf z \in \mathcal Z$ and every sample size: the data identify the centred kernel $\tilde\phi = \phi - \langle\phi, h\rangle$ and the combinations $\alpha + M_0\langle\phi, h\rangle$ and $\tau - q\langle\phi, h\rangle$, but not $\langle\phi, h\rangle$ itself.
\end{enumerate}
\end{proposition}

Three consequences frame what follows. First, the non-identification is structural, not statistical: by (iv) it holds at any sample size and cannot be broken by modelling the second moment, since in a mean--variance family the working variance is a function of the invariant $\boldsymbol\eta$ and the random-effects covariance does not involve $(\alpha,\tau,\phi)$. This consequence persists under temporal extension, where free period effects absorb the level regardless of rollout sequence (\S\ref{sec:temporal}). Only a design feature placing $s_0(\mathbf z)$ outside $\mathrm{col}\{X_0(\mathbf z)\}$ identifies the level (\S\ref{sec:design}). Second, by (iii) the exposure of an estimand to the unidentified level is measured by the single scalar $\ell(w,c)$; Table~\ref{tab:estimands} catalogues its value for the standard trial estimands, including $\ell = q \neq 0$ for the direct effect itself. Third, our results are statements on the linear-predictor scale. Outcome-scale contrasts such as risk differences are nonlinear functionals of $\boldsymbol\eta$ and inherit the non-identification through the link whenever they depend on the kernel level.

Estimands with $\ell(w,c) = 0$ are invariant to the unidentified level and are identified by randomisation alone. Distance contrasts of the marginalised response curve of \citet{Wang2025} are of this form. The curve itself is not and under the additive model their average marginalized effect at distance $d$ is $\phi(d)$, the estimand $(\Delta_d, 0)$ with $\ell = 1$, and its identification in their framework is supplied by the design rather than by level-invariance. Bernoulli assignment generates the allocation-to-allocation exposure variation that part (ii) shows is absent conditional on any realised allocation under a fixed margin. Their own treatment of completely randomised designs, where the marginal contrast loses its ceteris paribus interpretation and is recovered only asymptotically under local interference, is precisely this aliasing, and its resolution by local interference is the compact-support anchor of \S6, met from the design-based side. The remaining rows are the estimands that fall outside the invariant class and require either an anchor or a design that supplies level information.

Proposition~\ref{prop:level_aliasing} shows the constants are lost. For a kernel class $\mathcal H$, define the design null set $\mathcal N := \{\psi \in \mathcal H - \mathcal H : S[\psi](\mathbf z) \in \mathrm{col}(X_0(\mathbf z))\}$, the shape differences the realised design cannot distinguish from intercept and treatment adjustments. Then the following proposition shows when the shape is identified.

\begin{proposition}[The shape is identified]
\label{prop:shape_identification}
Say the design is \emph{contrast-complete} for $\mathcal H$ if $\mathcal N$ contains only functions constant $h$-a.e. Then any two triples generating the same linear predictor at $\mathbf z$ have kernels differing by a constant: the centred kernel $\tilde\phi = \phi - \langle\phi, h\rangle$ is identified, the equivalence classes are exactly the constant shifts $\{\phi + a: a \in \mathbb R\}$.
\end{proposition}

\section{Estimation and Bias}
\label{sec:bias}
\subsection{Working Model}
Our statistical model for the analysis of bias for observation $i = 1,...,N$ in cluster $k(i)$ is
\begin{equation}
 \label{eq:spatmodobs}
     y_i = \alpha + \tau z_{k(i)} +  S[\kappa]_i(\mathbf{z}) + \varepsilon_i, \hspace{1cm} \varepsilon \sim N(0,\Sigma)
 \end{equation}
where the kernel $\kappa$ ranges over a working class $\mathcal{K}$, a set of functions on $[0,D]$, with example families given below in \S\ref{sec:estimators}. By Propoisition \ref{prop:level_aliasing} we require $\mathcal{K} \cap \{constants\} = \{0\}$. The spatial covariance matrix is $\Sigma$, which is assumed stationary and isotropic, with $\Omega = \Sigma^{-1}$.

For this model we define the working analysis matrix as
\begin{equation*}
    X_{\mathcal K}(\mathbf z) = [\mathbf 1, \mathbf z, S[\psi_1](\mathbf z),...,S[\psi_J](\mathbf z)]
\end{equation*}
where $\mathbf z$ is excluded if there is no direct treatment column. And we further define the level-augmented working matrix as:
\begin{equation*}
    X_{\mathcal K}^+(\mathbf z) = [X_{\mathcal K}(\mathbf z), s_0(\mathbf z)]
\end{equation*}

\subsection{Anchors}
\label{sec:anchors}
Every estimand with $\ell(w,c) \neq 0$ references the counterfactual $Y_i(0,\mathbf{0})$. For many spillover kernels, all individuals in the study area may have some level of exposure to a source, and Propositions \ref{prop:level_aliasing} and \ref{prop:shape_identification} show that the data only provides information about the shape of the kernel and not its level in these cases. We define an `anchor' as the empirical stand-in for the untreated counterfactual. More formally, an \textit{anchor} is a pair $(A, a_0)$, where $A$ is a linear functional on kernels with $A(\mathbf{1})=1$ and $a_0 \in \mathbb{R}$ is its assumed value (throughout, $a_0 = 0$). The anchor selects from each equivalence class $\{\psi + a: a\in \mathbb{R}\}$ the unique representative satisfying $A(\psi) = a_0$. The \textit{anchoring error} at the truth is:
\begin{equation*}
    \bar{e}  := A(\phi)-a_0
\end{equation*}
the amount by which the true kernel violates the anchoring assumption. We also define the kernel error $e = \phi - \kappa$ and shape error $\tilde{e} = e - \bar{e}$.

\subsection{Estimators}\label{sec:estimators}
Every estimator we consider is the GLS fit, with weight $\Omega$ (estimated under C6), of $y$ on $X_{\mathcal K} = [\mathbf 1, \mathbf z, S[\psi_1], \ldots]$, profiled over $\rho$ in nonlinear cases; estimators differ only in the working class $\mathcal K$:
\begin{itemize}
    \item \textit{Parametric.} $\mathcal K = \{b\,\psi_p(\cdot;\rho) :     b \in \mathbb R, \rho \in \Theta\}$, a working kernel family of any     prescribed smoothness, e.g.\ exponential or Mat\'ern kernels, or
    Wendland's compactly supported constructions \citep{Gneiting2002}.
    \item \textit{Semiparametric.} Fix a truncation radius $R < D$ and order $p \ge 1$; $\mathcal K = \mathcal P_p\,\mathbf 1_{[0,R]}$ with     $\{\psi_j\}_{j=0}^{p}$ an orthonormal basis for $\mathcal P_p[0,R]$
    in $L^2(h)$.
    \item \textit{Traditional cluster trial.} $\mathcal K = \{0\}$: no spillover columns.
    \item \textit{Dose-response.} $\mathcal K$ parametric or
    semiparametric as above, with $\mathbf z$ omitted because units are not themselves treated; only estimands with $c = 0$ are available, and the level aliasing persists with the intercept in place of the period pair, $s_0 = M_0\mathbf 1$, at leverage $\ell = m(w)$.
\end{itemize}
The fitted element of $\mathcal K$ estimates a projection, not $\phi$. For the parametric family, let $\rho^*$ denote the pseudo-true lengthscale (C6) and define
\begin{equation*}
    b = \frac{\langle \tilde\phi,     \tilde\psi_p(\cdot;\rho^*)\rangle_h}{\lVert     \tilde\psi_p(\cdot;\rho^*)\rVert_h^2}, \qquad
    a = \langle \phi, h\rangle - b\,\langle \psi_p(\cdot;\rho^*),
    h\rangle,
\end{equation*}
with the analogous $L^2(h)$ projection coefficients in the linear case. The two coefficients have opposite statuses: $b$ is identified (Proposition~\ref{prop:shape_identification}), while $a$ is the level coordinate of the flat direction in Proposition~\ref{prop:level_aliasing}(ii) and is supplied by the estimator's anchor. Table~\ref{tab:anchors} lists, for each estimator, the induced anchor $A$, the anchored projection $\kappa$ (satisfying $A(\kappa) = 0$), and the resulting errors $\bar e = A(\phi)$ and $\tilde e = \tilde\phi - b\,\tilde\psi_p(\cdot;\rho^*)$. For a target (\ref{eq:linearf}) the plug-in estimator is 
\begin{equation*}
    \hat\theta \;=\; c\hat\tau + \langle \hat\kappa, w\rangle,
\end{equation*}
evaluable for any $w$ (with $\hat\kappa = 0$ for the traditional analysis, recovering the standard estimator).

\begin{table}[]
\small
    \centering
    \begin{tabularx}{\textwidth}{c|>{\raggedright\arraybackslash}Xcc}
    \toprule
    Estimator & Anchor functional $A(f)$ & $\bar{e}$ & $\tilde{e}$ \\
    \midrule
    Parametric     & Affine-projection intercept: $\bar{f} - b(f)\bar{f}(\rho)$ & $a$ & $r$\\
    Semi-parametric     & Tail mean: $\left( \int_{R}^{D}h(u) du\right)^{-1} \int_{R}^{D}h(u)f(u) du$ & mean $\phi$ on $(R, D]$ & $\sim r_m$ \\
    Traditional CRT & Between-cluster mean: $\langle f, h_b \rangle$ & $\bar{f}_b$ & $f -\bar{f}_b$ \\
    \bottomrule
    \end{tabularx}
    \caption{Three classes of estimator and their associated anchors and errors}
    \label{tab:anchors}
\end{table}

\subsection{Conditions}
We state a high-level condition we rely on for subsequent analyses. The remaining Conditions C2–C7 (provided in Supplementary Material) impose standard regularity conditions: geometry stabilises, the working precision is well-conditioned, covariance parameters are estimable, a CLT holds. 

Fix an estimand direction $(w,c)$, working class $\mathcal K = \mathrm{span}(\psi_1, \ldots,\psi_J)$, and weights $\Omega$, and suppose $X_{\mathcal K}(\mathbf z)^\top \Omega\, X_{\mathcal K}(\mathbf z)$ is nonsingular. For any $f$ with $S[f](\mathbf z)$ well defined, the \emph{projection functional} is
\begin{equation*}
\mathcal P_{N,w,c}(f;\mathbf z) := a_{w,c}^\top \{X_{\mathcal K}(\mathbf z)^\top\Omega X_{\mathcal K}(\mathbf z)\}^{-1} X_{\mathcal K}(\mathbf z)^\top\Omega\, S[f](\mathbf z), \qquad a_{w,c} = \bigl(0,\ c,\ \langle\psi_1,w\rangle,\ldots,\langle\psi_J,w\rangle\bigr)^\top,
\end{equation*}
and the \emph{effective level leverage} is
\begin{equation*}
L_{N,w,c}(\mathbf z) := m(w) - \mathcal P_{N,w,c}(\mathbf 1;\mathbf z).
\end{equation*}

$\mathcal P_{N,w,c}(f;\mathbf z)$ is the amount by which an omitted spillover component $f$ in the true mean is absorbed into the plug-in estimand by the GLS fit. 

\begin{condition}[Anchor-compatibility]
\label{cond:C4}
Fix an estimand direction $(w,c)$, a working class $\mathcal K$, weights $\Omega$, and an anchor $(A, a_0)$. Let $\mathcal R \subset \mathcal H - \mathcal H$ be the residual class: a linear class of candidate spillover misspecifications, equipped with norm $\|\cdot\|_{\mathcal R}$, containing the anchored error $\tilde e = e - A(e)\mathbf 1$ for every $\phi$ under consideration, and closed under centring $f \mapsto f - A(f)\mathbf 1$. Write $\mathcal R_0 = \{f \in \mathcal R : A(f) = 0\}$.

\begin{enumerate}
\item[(a)] \emph{(Nondegeneracy)} There is $c_0 > 0$ such that, with probability tending to one, $\lambda_{\min}\{\nu_N^{-1} X_{\mathcal K}(\mathbf z)^\top \Omega\, X_{\mathcal K}(\mathbf z)\} \ge c_0$, where $\nu_N$ is the information scale of the sampling regime ($\nu_N = K$ under increasing domain; $\nu_N = N$ under infill with observation-level noise).
\item[(b)] \emph{(Negligible shape projection)}
\begin{equation*}
\sup_{f \in \mathcal R_0,\, f \neq 0}\; \frac{\bigl|\mathcal P_{N,w,c}(f;\mathbf z)\bigr|}{\|f\|_{\mathcal R}}
\;=\; o_p(1).
\end{equation*}
\end{enumerate}
The design, working class, weights, and anchor are \emph{anchor-compatible} for $(w,c)$ if (a) and (b) hold.
\end{condition}

The constant direction is excluded deliberately in Condition \ref{cond:C4} as $\mathcal P_{N,w,c}(\mathbf 1;\mathbf z)$ requires no asymptotic condition. It is a computable design constant, equal to $-qc$ under (H1)--(H3) and to $m(w) - L_{N,w,c}(\mathbf z)$ in general, and its contribution to the bias is the exact level term of Theorem~\ref{thm:projection_bias}. For general $f \in \mathcal R$, decomposing $f = A(f)\mathbf 1 + \tilde f$ and applying (b) recovers $\mathcal P_{N,w,c}(f;\mathbf z) = \mathcal P_{N,w,c}(\mathbf 1;\mathbf z)A(f) + o_p(1)\|f\|_{\mathcal R}$.

Condition 1(b) is the estimation-level counterpart of contrast completeness (Proposition~\ref{prop:shape_identification}). The identification condition asserts that no anchored shape is \emph{exactly} aliased with the design column, it asserts that anchored shapes are asymptotically orthogonal, in the $\Omega$-projection sense, to the particular estimating direction $a_{w,c}$. Condition 1(a) is the quantitative form of the requirement that the working columns be well separated from the aliased subspace; it fails precisely when some $S[\psi_j]$ is nearly constant on the realised exposure distribution, the failure mode determined by $I_0$.

Both parts depend on the anchor only through the centring. Changing $A$ changes which representative of each equivalence class must project negligibly. A design compatible with a tail-mean anchor need not be compatible with a point anchor at $D$, since the two centrings differ by functions concentrated where the design may have sparse distance support.

\subsection{General bias of anchored estimators}
We now present the main bias result for anchored estimators. 

\begin{theorem}[Projection bias of anchored plug-ins]
\label{thm:projection_bias}
Conditional on the realised allocation $\mathbf z$,
\begin{equation*}
\mathbb E(\hat\theta \mid \mathbf z) - \theta = \mathcal P_{N,w,c}(e;\mathbf z) - \langle e, w\rangle = \underbrace{-\,L_{N,w,c}(\mathbf z)\,\bar e}_{\text{level}} \;\underbrace{-\;\langle \tilde e, w\rangle}_{\text{shape}} \;+\;\underbrace{\mathcal P_{N,w,c}(\tilde e;\mathbf z)}_{\text{leakage}} .
\end{equation*}
Under the canonical design (H1--H3), $L_{N,w,c}(\mathbf z) = \ell(w,c) = m(w) + qc$ exactly; if additionally Condition 1 holds, the leakage term is $o_p(1)$ and
\begin{equation*}
\mathbb E(\hat\theta) - \theta = -\,\ell(w,c)\,\bar e \;-\; \langle \tilde e, w\rangle \;+\; o_p(1).
\end{equation*}
\end{theorem}

Two invariances sharpen the reading of the level term. First, it is invariant to the link and the working weights. For any model in the class of Proposition~\ref{prop:level_aliasing}(iv) fitted by quasi-likelihood, the first-order bias admits the same decomposition with $\Omega$ replaced by the working precision at the fit. And, since $s_0(\mathbf z) \in \mathrm{col}\{X_0(\mathbf z)\}$ under (H1)--(H3) regardless of the metric, the projection of the constant direction, and hence the level term $-\ell(w,c)\bar e$, is unchanged. Only the shape and leakage terms are reweighted. Second, the level term is eliminated exactly when the level is estimated. If the design breaks the aliasing ($I_0 > 0$, Proposition~\ref{prop:level_information}) and the working model includes the column $s_0(\mathbf z)$ with plug-in weight $m(w)$, then $\mathcal P_{N,w,c}(\mathbf 1;\mathbf z) = m(w)$, $L_{N,w,c}(\mathbf z) = 0$, and the bias reduces to shape error and leakage alone. The choice between these two routes of assuming the level or estimating it is priced by Proposition~\ref{prop:level_information}(iii). 

\begin{corollary}[The traditional CRT analysis]
\label{cor:traditional_crt}
Under complete randomisation of $K_1$ of $K$ equal-size clusters, the difference-in-means estimator satisfies
\begin{equation*}
\mathbb E\bigl[\hat\tau_{\mathrm{crt}}\bigr] = \tau + (m_0 - q)\,\bar\phi_w - m_0\,\bar\phi_b = \tau_{\mathrm{cluster}} - m_0\,\bar\phi_b .
\end{equation*}
It is the plug-in of Theorem~\ref{thm:projection_bias} with empty working class and induced anchor $A(f) = \langle f, h_b\rangle$: for the target $\tau_{\mathrm{cluster}}$ its leakage exactly offsets its shape error, leaving pure level bias $-\ell(w_W,1)\bar\phi_b$ with $\ell(w_W,1) = m_0$; for $\tau_{DE}$ it is additionally biased by the absorbed within-cluster spillover $\langle\phi, w_W\rangle$. It is unbiased for $\tau_{\mathrm{cluster}}$ and hence $ATE$ if and only if the between-mean anchor is valid, $\bar\phi_b = 0$ as holds by design when all between-cluster distances exceed the kernel range.
\end{corollary}

Under (H1)---(H3), $S[\mathbf 1](\mathbf z)=M_0\mathbf 1-q\mathbf z$ so $\mathcal P_{N,w,c}(\mathbf z)=-qc$ and $L_{N,w,c}(\mathbf z)=m(w)+qc=\ell(w,c)$. The original level leverage $\ell(w,c)$ is therefore the effective level leverage in the canonical spaced-CRT setting. \citet{Leung2025} sharpens the fried-egg exclusion to units not surrounded by same-arm clusters, an analysis-stage form of the same anchor validation. Their observation that the classical version shifts the estimand to the cluster-interior subpopulation applies under unit-level heterogeneity, which the common-kernel model relegates to the exposure-misspecification caveat.

In designs that break the constant-kernel aliasing and include $S[\mathbf 1](\mathbf z)$ as an estimable level column, $\mathcal P_{N,w,c}(\mathbf z)=m(w)$ and the level term vanishes without requiring an external anchor. If none of these conditions is credible, estimands with nonzero effective level leverage remain sensitive to the unobserved level of the spillover kernel.
Fix reference constants $(M_0, q)$ and define the \emph{exposure perturbation}
\begin{equation*}
r(\mathbf z) := s_0(\mathbf z) - M_0\mathbf 1 + q\tilde{\mathbf z}.
\end{equation*}

\begin{proposition}[Level information and the price of not anchoring]
\label{prop:level_information}
Let $X_{\mathcal K}(\mathbf z)$ denote the anchored working design and $X_+(\mathbf z) = [X_{\mathcal K}(\mathbf z), s_0(\mathbf z)]$ its enrichment by the level column, with $\Omega = \Sigma^{-1}$. Define the \emph{level information}
\begin{equation*}
I_0(\mathbf z) := \min_{b}\,\{s_0 - X_{\mathcal K}(\mathbf z) b\}^\top \Omega \{s_0 - X_{\mathcal K}(\mathbf z) b\},
\end{equation*}
and suppose $I_0(\mathbf z) > 0$. Let $\hat\theta_-$ and $\hat\theta_+$ be the anchored and enriched GLS plug-ins for $\theta(w,c)$. Then, conditional on $\mathbf z$:
\begin{enumerate}
\item[(i)] \emph{(Exact variance decomposition)}
$\mathrm{Var}(\hat\theta_+ \mid \mathbf z) = \mathrm{Var}(\hat\theta_- \mid \mathbf z) + L_{N,w,c}(\mathbf z)^2 / I_0(\mathbf z)$.
\item[(ii)] \emph{(Near-aliasing bound)} If $s_0(\mathbf z) = M_0\mathbf 1 - q\tilde{\mathbf z} + r(\mathbf z)$, then since $\mathbf 1, \tilde{\mathbf z} \in \mathrm{col}(X_{\mathcal K})$, $I_0(\mathbf z) = \min_b \|r - X_{\mathcal K} b\|_\Omega^2 \le r(\mathbf z)^\top \Omega\, r(\mathbf z)$.
\item[(iii)] \emph{(Anchoring rule)} Up to shape terms, $\mathrm{MSE}(\hat\theta_-) - \mathrm{MSE}(\hat\theta_+) = L_{N,w,c}^2\,\{\bar e^{\,2} - I_0^{-1}\}$: anchoring dominates if and only if $|\bar e| < I_0(\mathbf z)^{-1/2}$.
\end{enumerate}
\end{proposition}

Proposition~\ref{prop:level_information} reduces level identification to a single question: how much $\Omega$-weighted variation does the constant-exposure vector $s_0(\mathbf z)$ retain after the working columns are partialled out? Writing $s_0(\mathbf z) = M_0\mathbf 1 - q\tilde{\mathbf z} + r(\mathbf z)$, the idealised part is annihilated by the projection, so
\begin{equation*}
I_0(\mathbf z) = \min_b\,\|r(\mathbf z) - X_{\mathcal K}(\mathbf z)b\|_\Omega^2
\;\le\; \|r(\mathbf z)\|_\Omega^2 .
\end{equation*}
All level information therefore resides in the exposure perturbation $r(\mathbf z)$, and only in the part of it that the working columns cannot reproduce. We define design choices generating peturbations in \S\ref{sec:design}.

The shape term is controlled by approximation only in the directions $w$ weights: for absolutely continuous targets $w(\mathrm du) = g(u)h(u)\, \mathrm du$, $|\langle\tilde e, w\rangle| \le \|\tilde e\|_h\|g\|_h$, while targets with atoms, such as $\theta(d)$, require pointwise control of $\tilde e$ at the evaluation distances (whence the continuity clause of C2). The leakage term, by contrast, is a property of the realised fit, design, weighting, and working basis, rather than of the target, and is $o_p(1)$ under Condition 1(b). 

\subsection{Inference and the sampling regime}
\label{sec:inference}
Conditional on $\mathbf z$, $\hat\theta = g_w^\top(\hat\tau, \hat\beta)^\top$ with $g_w = \bigl(c, \langle\psi_1, w\rangle, \ldots, \langle\psi_J, w\rangle\bigr)^\top$ is a fixed linear combination of GLS coefficients, so under C7 it is asymptotically normal about $\mathbb E(\hat\theta \mid \mathbf z)$ with variance $V_N(\mathbf z) = g_w^\top V(\mathbf z) g_w$, where $V(\mathbf z)$ is the GLS covariance of $(\hat\tau, \hat\beta)$; under C6 the same holds with estimated covariance parameters. The rate is governed by the information scale $\nu_N$ of Condition 1(a): under increasing domain ($K \to \infty$, bounded cluster sizes, summable correlations), $\nu_N = K$ and $V_N(\mathbf z) = O(K^{-1})$. Under infill of a bounded domain, any observation-level independent variation, including individual noise, a nugget, binomial or Poisson sampling, gives $X_{\mathcal K}^\top \Omega X_{\mathcal K} \asymp N$ and $V_N(\mathbf z) \to 0$. Only a nugget-free, purely spatially correlated process saturates at a positive limit \citep{ Zhang2005}; there C5 fails and the asymptotics of this section are not claimed.

Two things about this otherwise standard theory are not standard. First, the calibration: the interval is centred at the anchored target $\mathbb E(\hat\theta \mid \mathbf z)$, and covers the causal estimand $\theta$ only when the bias of Theorem~\ref{thm:projection_bias} vanishes. Second, its interaction with that bias: the level term is invariant to the regime, the nugget, and the outcome model, so wherever the variance vanishes the mean squared error does not.

\begin{corollary}[The bias floor]
\label{cor:bias_floor}
Fix an estimand with $\ell(w,c) \neq 0$ under the canonical design (H1--H3) and an anchored working model satisfying Condition 1. Along any sequence of designs with $V_N(\mathbf z) \to 0$, increasing domain, or infill with observation-level noise:
\begin{equation*}
\mathrm{MSE}(\hat\theta \mid \mathbf z) \;\longrightarrow\; \bigl\{\ell(w,c)\,\bar e + \langle \tilde e, w\rangle\bigr\}^2,
\end{equation*}
equal to $\ell(w,c)^2\bar e^{\,2}$ when the working class spans the centred kernel. Additional data of the same design sharpens the estimate of the anchored target but cannot move it toward the causal estimand.
\end{corollary}

\subsection{Summary}\label{sec:summary}
A single design-computable scalar, the level leverage, measures an estimand's exposure to the unidentified level of the spillover function: it gives the direction of non-identifiability, the exact first-order bias of any anchored estimator, the variance premium for estimating the level instead, and the MSE floor that no amount of data can remove.

\begin{tcolorbox}[colback=white, colframe=black, boxrule=0.5pt,
  sharp corners, left=6pt, right=6pt, top=6pt, bottom=6pt,
  title={The level-leverage identity}, coltitle=black,
  colbacktitle=white, fonttitle=\bfseries,
  attach title to upper={\par\smallskip}]
Fix an estimand $(w,c)$, a design, a working class, weight $\Omega$ and anchor, and write $L = L_{N,w,c}(\mathbf z)$; under (H1)--(H3), $L = \ell(w,c) = m(w) + qc$ exactly. Then:
\begin{enumerate}
\item[(i)] \emph{Identification}
(Proposition~\ref{prop:level_aliasing}(iii); exact): the reparameterisation $\phi \mapsto \phi + a$ moves $\theta$ by $a\,\ell(w,c)$; the estimand is invariant to the unidentified level if and only if $\ell(w,c) = 0$.
\item[(ii)] \emph{Bias} (Theorem~\ref{thm:projection_bias}; up to shape terms): the anchored plug-in satisfies $\mathbb E(\hat\theta \mid \mathbf z) - \theta = -L\,\bar e$.
\item[(iii)] \emph{Variance}
(Proposition~\ref{prop:level_information}(i); exact, requiring $I_0(\mathbf z) > 0$): estimating the level rather than anchoring it costs $L^2/I_0(\mathbf z)$ in conditional variance.
\item[(iv)] \emph{Risk} (Corollary~\ref{cor:bias_floor}; canonical design, Condition 1): if $\ell \ne 0$ and $V_N \to 0$, $\mathrm{MSE}(\hat\theta) \to \ell^2\bar e^{\,2}$ under a well-specified working class.
\end{enumerate}
\end{tcolorbox}

\section{Design for spatial estimands}
\label{sec:design} 
The preceding results turn design choice into a well-posed problem. Two tasks must be separated. The first is \emph{level identification}: pinning the additive level of the kernel, which fixed-margin randomisation leaves aliased (Proposition \ref{prop:level_aliasing}). The second is \emph{shape efficiency}: estimating the centred spillover-response kernel precisely in the part of the distance distribution that the target estimand weights.

\subsection{Level information and identification}
Each augmentation in Table \ref{tab:taxonomy} modifies exactly one primitive of the canonical design: the exposure sets $\mathcal T_i$ (H1), the exclusion sets $\mathcal A_i$ (H2), or the allocation support $\mathcal Z$ (H3). Writing $c_i(\mathbf z) = \sum_{t\in\mathcal T_i} z_{\iota(t)}$ for the treated source count of unit $i$, so that $s_{0,i}(\mathbf z) = c_i(\mathbf z)$, every perturbation takes the form $r_i(\mathbf z) = c_i(\mathbf z) - M_0 + q z_{k(i)}$. The designs differ only in what generates variation in $c_i$:
\begin{itemize}
    \item \emph{Margin  variation} enlarges $\mathcal Z$ beyond the fixed-margin set (e.g.\ Bernoulli assignment): with H1--H2 intact, $c_i = M(\mathbf z) - qz_{k(i)}$ still, so $r(\mathbf z) = \{M(\mathbf z)-M_0\}\mathbf 1 \in \mathrm{col}(X_0)$ at every realised allocation. 
    \item \emph{Sentinel units} are a set $\mathcal S$ of $n_s$ units with $\mathcal T_i = \varnothing$, so $c_i = 0$ on $\mathcal S$.
    \item \emph{Baseline periods} extend the index to unit--periods with rollout between periods, so $\mathcal T_{i,0} = \varnothing$ and $c_i = 0$ on all pre-rollout observations---sentinel units in time  (\S\ref{sec:temporal}); the perturbation is nonzero only once the trend is restricted, since free period effects return it to $\mathrm{col}(X_0)$.
    \item \emph{Randomised saturation} localises the exposure sets and randomises within-cluster treatment proportions in a first stage, so $c_i$ varies within allocation by construction.
    \item \emph{Truncated range} sets $\mathcal T_i = \{t : d_{i,t} \le D_{\max}\}$ through the working model; retaining H3, $r_i = -b_i(\mathbf z)$ where $b_i$ counts treated sources beyond range, and variation is generated by boundary geometry.
    \item \emph{Heterogeneous exclusion} relaxes H2 to $|\mathcal A_i| = q_{k(i)}$. 
    \item \emph{Buffers} leave H1--H3 intact and instead push between-cluster distances beyond the kernel's effective support.
\end{itemize}

\begin{table}
\centering\small
\begin{tabularx}{\textwidth}{lc>{\raggedright\arraybackslash}X>{\raggedright\arraybackslash}X>{\raggedright\arraybackslash}X}
\toprule
Design feature & Relaxes & $r_i(\mathbf z)$ & $I_0$ & Identifying assumption \\
\midrule
Fixed-margin CRT & --- & $0$ & $0$ in every regime & --- (anchor required) \\
Margin variation & H3 & $M(\mathbf z)-M_0$; $\ r\propto\mathbf 1$ & $0$ at every realised $\mathbf z$ & --- \\
Sentinel units & H1 on $\mathcal S$ & $-(M_0-qz_{k(i)})\mathbf 1\{i\in\mathcal S\}$ & $\asymp n_sM_0^2\times{}$sentinel $\Omega$-leverage & mean exchangeability of sentinel and exposed units \\
Baseline, trend restricted & adds $p=0$ & $-M_0\mathbf 1\{p=0\}$ & $\asymp N_{\mathrm{pre}}M_0^2\times{}$period-contrast $\Omega$-leverage; $0$ under free period effects & no secular shock concurrent with rollout
(pre-trends testable) \\
Randomised saturation & H1, H3 & $c_i(\mathbf z)-M_0+qz_{k(i)}$ & $\asymp$ within-allocation $\operatorname{Var}(c_i)$ & locality of $\mathcal T_i$ correct; exogeneity of $c_i$ enforced by randomisation \\
Truncated kernel range & H1 & $-b_i(\mathbf z)$ & $\asymp$ boundary unit count ${}\times\operatorname{Var}(b_i)$ & range correctly specified; no systematic edge effects \\
Heterogeneous exclusion & H2 & $(q-q_{k(i)})z_{k(i)}$ & $\asymp\operatorname{Var}_k(q_k)\times{}$treated units & cluster size non-informative \\
Buffer / spacing & --- & $0$ & $0$ & --- (validates anchor: $\bar\phi_b\to0$) \\
\bottomrule
\end{tabularx}
\caption{Design features classified by the canonical-design primitive they relax and the exposure perturbation  $r(\mathbf z) = s_0(\mathbf z) - M_0\mathbf 1 + q\mathbf z$ they generate. }
\label{tab:taxonomy}
\end{table}

Three remarks govern the use of Table~\ref{tab:taxonomy}. First, compact support of the kernel, $\phi(d) = 0$ for $d > R$, with design distance support beyond $R$, is a single assumption spendable on either route: as level information, generating the truncation and sentinel rows without pure controls by design, or as an anchor, rendering the tail mean exactly valid, $\bar e = 0$. When it holds the anchor route dominates, avoiding the variance premium $L^2/I_0$ of Proposition~\ref{prop:level_information}(i) at no cost in bias. It is partially testable: the identified centred kernel
(Proposition~\ref{prop:shape_identification}) must be constant on $(R, D]$, but the value of that constant is precisely $\bar e$, so the design can refute compact support but never confirm it, which licenses estimating $R$ from the identified shape, as in the contamination-range analyses of \citet{Multerer2021, AnayaIzquierdo2021, Watson2025}, while making explicit which component is data-driven and which is assumption.

Second, $I_0$ retains only variation in $r(\mathbf z)$ that the columns $S[\psi_1],\ldots,S[\psi_J]$ cannot mimic, so effective features are rough relative to $\mathcal K$; in practice, interleaving sentinels within the study region both weakens their exchangeability assumption and raises their $\Omega$-leverage, whereas the treatment-by-cluster-size interaction exploited by heterogeneous exclusion is the direction confounded by informative cluster size \citep{Kahan2023}, and edge deficits convert from information into bias under range misspecification. Only randomised saturation carries an assumption the design itself
enforces; the two-stage designs of \citet{Leung2025} are of this form. 

Third, the design choice for a level-dependent estimand is the anchoring rule of Proposition~\ref{prop:level_information}(iii): engineer $I_0$ with $\sigma/\sqrt{I_0} < |\bar e|$ under credible assumptions, or validate an anchor by design (buffers, Corollary~\ref{cor:traditional_crt}) and accept $\bar e \approx 0$ as an assumption. Both routes are priced by the same leverage $L_{N,w,c}$, and both prices are computable before unblinding.

\subsection{Temporal augmentation}\label{sec:temporal}
Baseline observation is a common augmentation for standard trial designs and the framework absorbs it without new machinery. The temporal mean model enters as columns of $X_0$. Extend the index to unit-periods $(i,p)$, $p \in \{0,1\}$, with rollout  between periods, so $\mathcal T_{i,0} = \varnothing$ and pre-rollout exposure is structurally zero. Write $\zeta_{(i,p)} = z_{k(i)}\mathbf 1\{p=1\}$ for the operative treatment and retain the static kernel (time-varying kernels are deferred to the discussion). A kernel shift $\phi \mapsto \phi + a$ moves no baseline mean; it moves follow-up means through $s_0(\mathbf z) = M_0\mathbf 1_{\{p=1\}} - q\boldsymbol\zeta$. With a free period effect, $s_0 \in \operatorname{col}(X_0)$ at every realised allocation, and Proposition~\ref{prop:level_aliasing}(i) applies verbatim: the flat direction $(\alpha_1 - aM_0, \tau + aq, \phi + a)$ absorbs the level into the period effect. The conclusion extends to any number of periods and any rollout sequence: under H1--H2 the within-period margin is constant, so period effects absorb the exposure wholesale, and the standard time-adjusted stepped-wedge model \citep{Hussey2007} is level-aliased under global exposure. Rollout varies the margin across periods but never within them.

Level information is bought by restricting the trend. Under a common-mean restriction the perturbation is $r(\mathbf z) = -M_0\mathbf 1_{\{p=0\}}$ and the  $N_{\mathrm{pre}}$ baseline observations are sentinel units in time, in the out-of-cluster form, with $I_0 \asymp N_{\mathrm{pre}} M_0^2 \times{}$period-contrast $\Omega$-leverage. The leverage behaves as for spatial sentinels since within-unit correlation raises it, while a period-level variance component $\sigma_\pi^2$ discounts it, interpolating between the independence value and re-aliasing as $\sigma_\pi^2 \to \infty$, the random-effects limit of the free period effect. The price of the restriction is an anchoring error, exactly. If the true drift is $\delta$, the identity $\mathbf 1_{\{p=1\}} = (s_0 + q\boldsymbol\zeta)/M_0$ decomposes the misspecification without remainder,
\begin{equation*}
  \delta\,\mathbf 1_{\{p=1\}}   = \frac{\delta}{M_0}\,s_0 + \frac{q\delta}{M_0}\,\boldsymbol\zeta,
  \qquad\text{whence}\qquad
  \operatorname{bias}(\hat\theta) = \ell(w,c)\,\frac{\delta}{M_0},
\end{equation*}
with no shape contamination for any $\Omega$. The anchoring rule of Proposition~\ref{prop:level_information}(iii) therefore extends verbatim and the choice is between extrapolating to zero exposure along the kernel ($|\bar e|$) and along the trend ($|\delta|/M_0$), each priced by the same leverage. Level-loaded estimands reference $Y(\mathbf 0)$, which the intervention preiod of the study never observes. A baseline period observes it at another time, and the trend restriction transports it. Like compact support, the restriction is refutable but not confirmable, with two or more pre-rollout periods the trend model is testable on exposure-free data, but a shock concurrent with rollout, which is the component aliased with the level, is not.

\subsection{Design stage metrics}
For a target estimand $(w,c)$, a linear working class $\mathcal K=\mathrm{span}(\psi_1,\ldots,\psi_J)$, a candidate geometry, and a working precision $\Omega$, three quantities are computable at the design stage.
 
\paragraph{(i) Variance.} The plug-in estimator $\hat\theta=c\hat\tau+\sum_j\hat\beta_j\langle\psi_j,w\rangle$ variance is $V_N(\mathbf z)$. Minimising the variance over geometries is a $c$-optimal design problem with target direction $g_w$; different estimands give different $g_w$, so no design is efficient for all of them, and the correlated-observation $c$-optimality algorithms of \citet{Watson2023b,Watson2023} apply directly.
 
\paragraph{(ii) Leakage.} For an omitted anchor-centred shape $f$ (with $A(f)=0$) the projection $\mathcal P_{N,w,c}(f;\mathbf z)$ is the bias that the realised geometry and randomisation inject into $\hat\theta$ beyond the genuine target-shape error. For a candidate design we can evaluate leakage directly . Let $\mathcal F$ be a prespecified library of anchor-centred residual kernels representing scientifically plausible departures from the working class. Define
\begin{equation}
\label{eq:leakdiag}
\Lambda_{\mathcal F,w,c}(\mathbf z)\;:=\; \max_{\;f\in\mathcal F,\;A(f)=0\;} \frac{\bigl|\mathcal P_{N,w,c}(f;\mathbf z)\bigr|}{\|f\|_h},
\end{equation}
This quantity is computable before outcomes are observed. A design with small $\Lambda_{\mathcal F,w,c}$ is locally robust to the omitted shapes in $\mathcal F$. A design with large $\Lambda_{\mathcal F,w,c}$ allows those omitted shapes to be absorbed into the plug-in estimand. Under homogeneous exclusion, fixed-margin randomisation, the constant direction has exact leakage $\mathcal P_{N,w,c}(f;\mathbf z)=-qc$. 
 
\paragraph{(iii) Level information.} The third diagnostic is the level information $I_0(\mathbf z)$ of Proposition~\ref{prop:level_information}: the $\Omega$-weighted variation in $s_0(\mathbf z)$ that the working columns cannot reproduce, computable from the candidate geometry, randomisation, and working model before any outcome is observed. Under the canonical design $I_0 = 0$ identically (Proposition~\ref{prop:level_aliasing}(ii)), and no rescaling of inter-cluster spacing changes this; positive $I_0$ requires a design feature generating residual variation in $s_0$ (Table \ref{tab:taxonomy}). When $I_0 > 0$ and the level column is fitted, $\mathrm{SE}(\hat\beta_0 \mid \mathbf z) = \sigma/\sqrt{I_0}$, and the anchoring rule of Proposition~\ref{prop:level_information}(iii) becomes a design criterion: estimate the level rather than anchor if and only if 
\begin{equation*} 
    \sigma/\sqrt{I_0(\mathbf z)} \;<\; |\bar e|, 
\end{equation*}
the plausible anchoring error. Because the leverage $\ell(w,c)^2$ multiplies both sides of the comparison, the rule is \emph{estimand-independent}: one computation of $I_0$ against one elicited bound on $|\bar e|$ settles the anchor-or-identify decision for every level-dependent estimand in Table~\ref{tab:estimands} simultaneously. \citet{Leung2025} elicits the analogous decay bound from transmission models and dispersal data, and prices it into a bias-aware interval, which is the anchoring rule executed at the inference stage rather than the design stage.

\begin{figure}
    \centering
    \includegraphics[width=\linewidth]{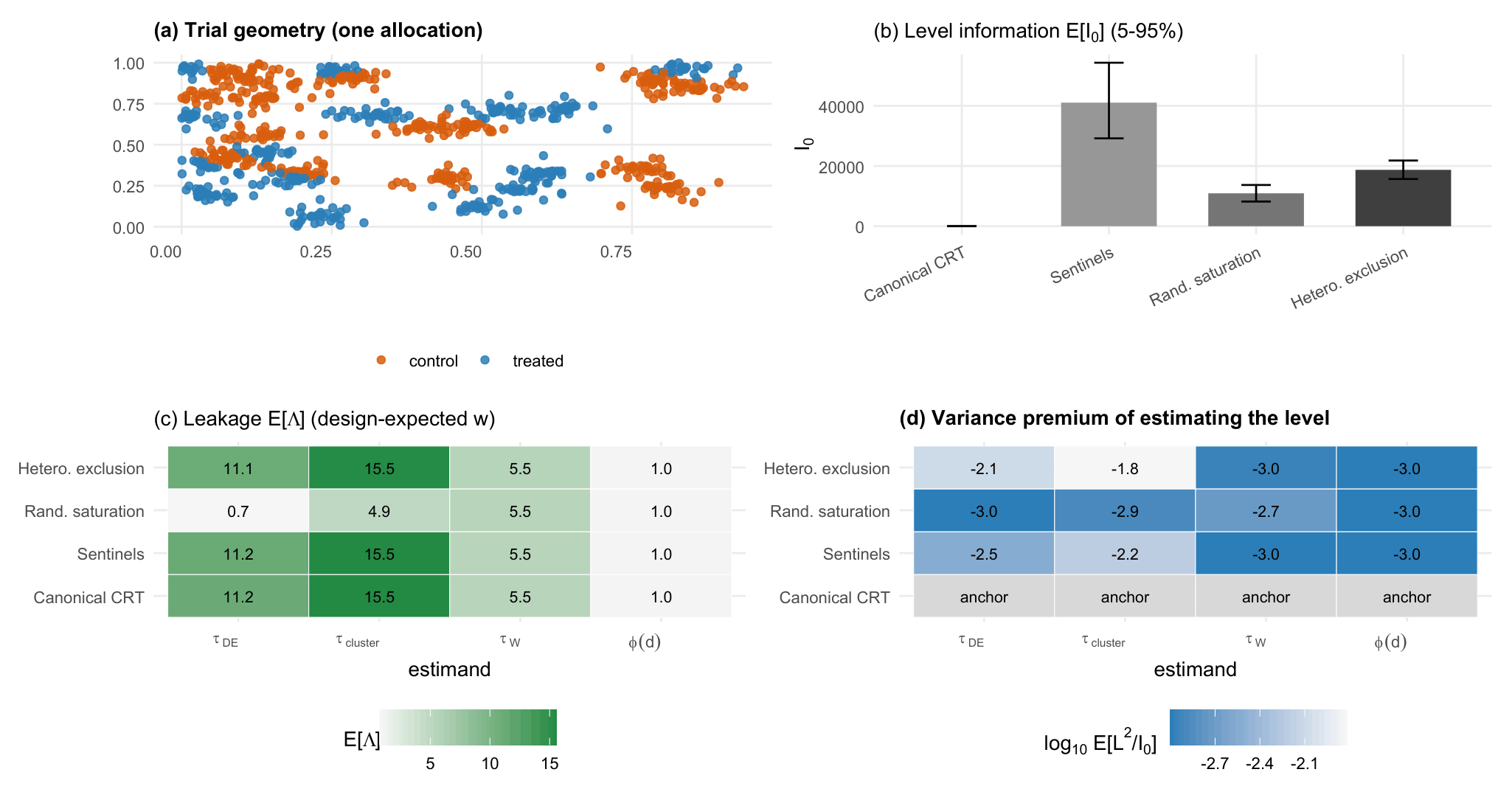}
    \caption{Example design stage diagnostics for a cluster randomised trial design}
    \label{fig:diagnostics}
\end{figure}

Figure \ref{fig:diagnostics} illustrates the design-stage diagnostics for the spillover estimators using the geometry simulation of \S\ref{sec:simulation} with $K=30$, $n=30$, and where sources are treated units. The canonical cluster trial has $I_0 = 0$ since its constant-exposure vector lies in the span of the intercept and treatment columns, so the kernel level is unidentified at any sample size. Each augmentation generates residual exposure variation that identifies the level, with randomised saturation and sentinel units the most informative and heterogeneous exclusion the weakest. The leakage diagnostic $\Lambda$ shows the worst-case absorption of an omitted spillover shape into the estimator, per unit norm of that shape. Larger values flag estimands whose realised bias is most sensitive to getting the kernel shape wrong, here the direct effect and pointwise dose-response. For a design with level information $I_0$, estimating the level is preferred to anchoring when the plausible anchoring error $|\bar e|$ exceeds $\mathrm{SE}(\hat\beta_0) = 1/\sqrt{I_0}$, shown across candidate values of $|\bar e|$. The canonical trial always anchors; designs with sufficient $I_0$ cross to estimation as either the identifying information or the tolerated anchoring error grows.

\section{Simulation study}
\label{sec:simulation}
The closed-form illustration isolates the conditional bias; we complement it with a Monte Carlo study confirming that the estimators behave under noise as the asymptotic theory predicts. Cluster centroids are placed on the unit square and $n$ units are scattered about each and assigned to the nearest centroid, giving $K$ balanced-randomised clusters. The geometry is fixed within a scenario while the allocation $\mathbf z$ and the outcomes are redrawn across replications. Outcomes are Gaussian with mean $\alpha+\tau  z_{k(i)}+S[\phi]_i(\mathbf z)$, a spatial Gaussian-process random effect, and an independent residual. We take the true kernel $\phi=\delta\,\kappa_0$ with $\kappa_0$ ranging over Mat\'ern shapes of varying smoothness and a compactly supported Wendland shape, spanning the smooth and compact cases. We include two scenarios: `high' has $\delta = 0.03$ with lengthscale of 0.1, and `low' sets $\delta = 0.003$ and lengthscale to 0.05. In all scenarios we set $\tau = 0.1$ and use a Matern spatial correlation function for the Gaussian process with smoothness 1.5, marginal variance 0.1 and lengthscale of 0.3.

We fit the three estimator classes of  \S\ref{sec:bias}: the traditional cluster-level analysis ($\mathcal K=\{0\}$), the semiparametric Chebyshev bases at two orders, and a parametric working kernel (Wendland), each by GLS with the Gaussian-process covariance approximated by a Hilbert-space expansion \citep{Solin2020} and the lengthscale profiled (using R package \textit{glmmrBase}, see \citet{Watson2026}). We do not include parametric kernels with non-linear parameters (e.g. Gaussian) for simplicity, to avoid the issue of generating intervals for profiled non-linear parameters. For every fit we form the plug-in $\hat\theta=c\hat\tau+\langle\hat\kappa,w\rangle$ and its model-based standard error $g_w^\top V(\mathbf z)\,g_w$ for each estimand of Table~\ref{tab:estimands}, including a dose-response value $\theta(d)$ and the zero-leverage contrast $\theta(d)-\theta(d')$.

\begin{figure}
    \centering
    \includegraphics[width=0.9\linewidth]{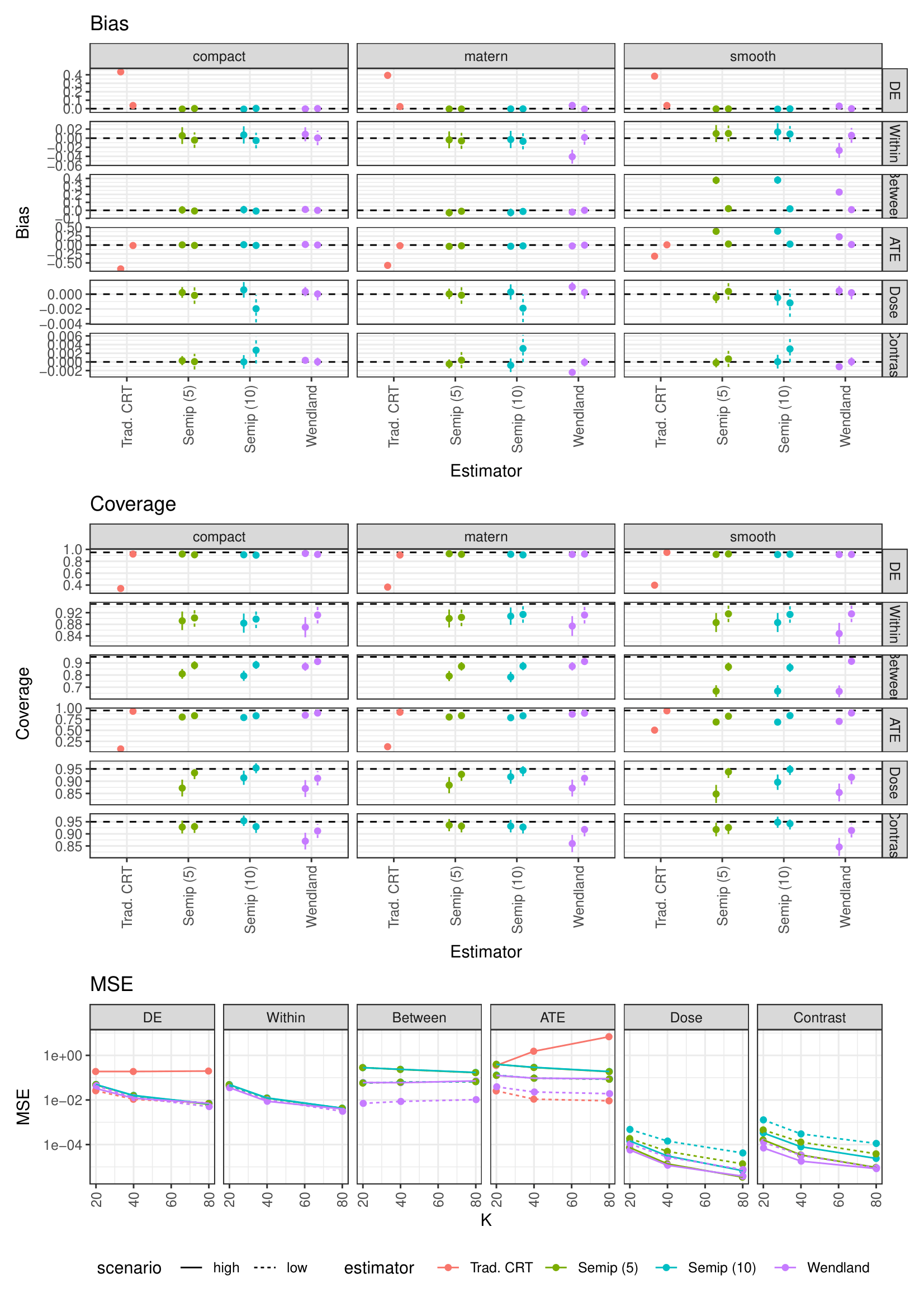}
    \caption{Simulation study results showing bias (top panel), coverage (middle panel) and mean squared error (bottom) panel}
    \label{fig:bias}
\end{figure}

Three findings bear out the theory shown in Figure \ref{fig:bias}. First, the results show that realised bias follows Theorem~\ref{thm:projection_bias}: for level-loaded estimands ($\ell(w,c)\neq0$) without a valid anchor (Matern, smooth) the flexible estimators are biased with bias increasing in the high spillover scenario. A richer kernel removes shape error but not the anchoring term $\ell(w,c)\bar e$, whereas the zero-leverage contrast and valid anchor effects (compact) are recovered with negligible bias by every spillover-aware estimator. Second, as $K$ grows the plug-in variance diminishes and the mean squared error of level-loaded estimands approaches the floor $\ell(w,c)^2\bar e^2$, while zero-leverage estimands are consistent. Third, the model-based standard errors are generally well calibrated for some estimands (dose-response, contrast, direct effect) whereas other estimators exhibit under-coverage resulting from the estimator bias and possible small sample issues.

\section{Conclusions}
The framework locates the existing approaches to spatial interference as points in a common space rather than competing methods. The estimands of \citet{Wang2025} reference the design's ambient exposure distribution. They  identify the level unconditionally through the allocation-to-allocation exposure variation that Bernoulli assignment supplies. This is the margin-variation route of Table~\ref{tab:taxonomy}, exercised without conditioning on the realised allocation, while distance contrasts of
their marginalised response curve sit in the level-invariant class $\ell = 0$ and incur no such cost. \citet{Leung2025} adopts a tail anchor with an explicitly elicited error bound and prices it into inference, the anchoring rule of Proposition~\ref{prop:level_information}(iii) executed at the analysis stage. Their well-surrounded exclusion is analysis-stage anchor validation, sharpening the buffer or fried egg designs of \citet{Hayes2009}, and their randomised-saturation design occupies the one taxonomy row whose identifying assumption the design itself enforces. \citet{Watson2025} takes the anchor-validation route in its limiting form, compact support with $\bar e = 0$ assumed exactly. 

Methods for spatially structured estimands imply that the traditional cluster randomised trial design may be estimating a non-transportable effect of less policy relevance than the ATE or direct effect. However, in cases where the desired estimand is the ATE and there is not ``too much'' spillover between clusters, the traditional CRT design has lower MSE. The efficiency cost of estimating the spillover may result in worse overall mean squared error. However, where interest is broader than a single ATE, or where spillover cannot be assumed away, the framework presented here suggests multiple avenues for modifying the design. However, we also note that more work is required to maximise the efficiency of such designs. For example, complete balanced randomisation schemes with some of the aforementioned features may have some very inefficient realisations and a restricted randomisation scheme can exclude `unstable' allocations to boost power. Where the trialist has a larger pool of study locations to generate a sub-sample from, there may be further potential gains in efficiency from optimal sample selection. 

A temporal dimension may also change the identification problem. We showed that identification from baseline observation still requires restrictions on temporal trend. However, a complete treatment of temporal effects may relax the static kernel and introduce carry-over effects. \citet{Mukaigawara2025} define estimands as contrasts between stochastic intervention strategies on a treatment point process, with interference and carryover left arbitrary. 

Our exact results are computed within the linear exposure mapping, and two of their components fare differently under its misspecification. The extrapolation argument above is model-free. Level-loaded estimands require an assumption bridging to the unobserved all-control state under any exposure structure, though outside the linear model that assumption is no longer a single scalar and the closed-form prices, $\ell(w,c)$, $I_0$, the contamination constant of Corollary~\ref{cor:traditional_crt}, do not survive as stated. The contrast class does survive, from the other direction. Under approximate neighbourhood interference, distance-marginalised contrasts remain estimable without a correctly specified exposure model \citep{Svje2021, Svje2024}, and these are precisely the $\ell(w,c) = 0$ estimands that our framework identifies as free of the extrapolation. 

The spatial paradigm requires more information at the design stage than a traditional approach. The most important component is the distance kernel or at least an effective range $R$ or lengthscale $\rho$. The spillover function may be difficult to estimate, particularly for new technologies, and will likely depend on the follow-up time of the study. At the design stage, a reasonable spillover kernel can be chosen for analyses since the shape and projection errors can be estimated for a wide range of kernels. Mathematical models of the spillover mechanisms may provide strong prior evidence to support choices of spillover kernels and length scale parameters. For example, a Gaussian kernel corresponds to a diffusion model \citep{Smith2026} and there exist mechanistic epidemiological models for many infectious diseases and interventions (e.g. \citet{Hancock2024} for gene drives for vector control) to estimate lengthscale parameters. For sample size analysis, some information on spatial correlations are also required. 

Spillover between clusters in a cluster randomised trial may imply several different policy-relevant estimands. Excluding spillover to satisfy SUTVA does not necessarily protect the trial and researchers, but potentially leads to untestable assumptions and non-transportable results. 





\bibliographystyle{abbrvnat}
\bibliography{adapt}

\end{document}